\documentclass[aps,prb,reprint,superscriptaddress, showpacs]{revtex4-1}

\usepackage{graphicx}
\usepackage{amsmath}
\usepackage{amssymb}
\usepackage{color}

\begin{document}


\title{Field induced quantum-Hall ferromagnetism in suspended bilayer graphene}



\author{H.~J.~van Elferen}  \email{helferen@science.ru.nl}
\affiliation{High Field Magnet Laboratory and Institute of Molecules and Materials, Radboud University Nijmegen, Toernooiveld 7, 6525 ED Nijmegen, The Netherlands.}
\author{A.~Veligura}  \email{a.veligura@rug.nl}
\affiliation{Physics of Nanodevices, Zernike Institute for Advanced Materials,
University of Groningen, Nijenborgh 4, 9747 AG Groningen, The Netherlands.}
\author{E.~V.~Kurganova}
\affiliation{High Field Magnet Laboratory and Institute of Molecules and Materials, Radboud University Nijmegen, Toernooiveld 7, 6525 ED Nijmegen, The Netherlands.}
\author{U.~Zeitler}
\affiliation{High Field Magnet Laboratory and Institute of Molecules and Materials, Radboud University Nijmegen, Toernooiveld 7, 6525 ED Nijmegen, The Netherlands.}
\author{J.~C.~Maan}
\affiliation{High Field Magnet Laboratory and Institute of Molecules and Materials, Radboud University Nijmegen, Toernooiveld 7, 6525 ED Nijmegen, The Netherlands.}
\author{N.~Tombros}
\affiliation{Physics of Nanodevices, Zernike Institute for Advanced Materials,
University of Groningen, Nijenborgh 4, 9747 AG Groningen, The Netherlands.}
\author{I.J. Vera-Marun}
\affiliation{Physics of Nanodevices, Zernike Institute for Advanced Materials,
University of Groningen, Nijenborgh 4, 9747 AG Groningen, The Netherlands.}
\author{B.~J.~van Wees}
\affiliation{Physics of Nanodevices, Zernike Institute for Advanced Materials,
University of Groningen, Nijenborgh 4, 9747 AG Groningen, The Netherlands.}

\begin{abstract}
We have measured the magneto-resistance of freely suspended high-mobility bilayer graphene. For magnetic fields $B>1$~T we observe the opening of a field induced gap at the charge neutrality point characterized by a diverging resistance. For higher fields the eight-fold degenerated lowest Landau level lifts completely. Both the sequence of this symmetry breaking and the strong transition of the gap-size point to a ferromagnetic nature of the insulating phase developing at the charge neutrality point.\\
\end{abstract}



\maketitle



\section{Introduction}
The unique electronic properties of monolayer and bilayer graphene makes them promising candidates for future applications in nanotechnology. Though (bilayer) graphene on a SiO$_2$-substrate can show a mobility up to 2~m$^2$/Vs,\cite{Geim1} much cleaner and higher mobility samples are necessary in order to investigate its intrinsic properties, and, in particular, electron interaction effects. Mobilities exceeding 10~m$^2$/Vs can be obtained by removing the SiO$_2$ substrate underneath the graphene\cite{Bolotin1,Du1} or by depositing graphene on a boron nitride crystal.\cite{Dean1}
These high-mobility samples display new interaction-induced phenomena such as a fractional quantum Hall effect,\cite{Dean2, Bolotin2, Du2} broken-symmetry states,\cite{Feldman1} a magnetic-field induced insulating phase,\cite{Feldman1} and quantized conductance at zero magnetic field.\cite{Tombros2}\footnotetext{check list mobilities on end}

In the two-dimensional electron system of bilayer graphene (BLG) the application of a perpendicular magnetic field results into an unconventional integer quantum Hall effect with plateaus at filling factors $\nu$ = $\pm 4, \pm 8, \pm 12$, ...\cite{Novoselov_UQHE} The lowest Landau level is eight-fold degenerate, owing to spin, valley and layer-index degrees of freedom. In standard BLG samples deposited on SiO$_2$, magnetic fields around 10~T are required to observe fully quantized plateaus and the eight-fold degeneracy of the lowest Landau level is only lifted for the highest quality samples at magnetic fields exceeding 20~T.\cite{Zhao1}
At 0~T the density of states in BLG does not vanish at the charge neutrality point, in contrast to single layer graphene, therefore, even arbitrarily weak interaction between charge from conduction and valence band states will trigger excitonic instabilities which causes a variety of gapped states.\cite{Nandkishore1,Nandkishore2,Zhang2,Jung1,Martin1}

In this paper, we present two-terminal magnetotransport experiments in suspended BLG at temperatures ranging from 1.3~K to 4.2~K and magnetic fields up to 30~T. We observe a sudden gap opening at the CNP already for $B \geq$~1~T and the appearance of broken-symmetry states at filling factors $\nu = \pm 1, \pm 2, \pm 3$ for higher fields. Detailed investigation of the energy gap at filling factor $\nu$ = 0 reveals an exchange-interaction driven linear scaling at low magnetic fields, in agreement with earlier reported results.\cite{Feldman1} At high fields we observe the cross-over to a much smaller gap. This high field transition and the appearance of broken symmetry states at $\nu$ = 1, 2, 3 are consistent with the formation of a quantum Hall ferromagnetic state.\cite{Barlas1,Nandkishore2}

\section{Experimental background}

We have prepared a suspended BLG sample using an acid free method.\cite{Tombros1} Following standard techniques,~\cite{Novoselov_firstgraphene} we first exfoliated flakes from highly oriented pyrolytic graphite (HOPG) and deposited them on a Si/SiO$_2$ substrate covered with a 1.15~$\mu$m thick LOR-A resist layer. Bilayer flakes were then identified by their optical contrast.~\cite{Blake1}
Subsequently, two electron beam lithography steps were performed in order to contact the flakes with Ti-Au contacts and to remove part of the LOR-A below the graphene flakes. The resulting device is freely suspended across a trench formed in the LOR-A with two metallic contacts on each side, see inset of Fig.~\ref{Fig1}.

Carriers in the BLG sheet can be induced by applying a back-gate voltage $V_G$ on the highly $n$-doped Si wafer. The geometrical gate capacitance is given by a combination of the vacuum gap (1.15~$\mu$m) and SiO$_2$ substrate (0.5~$\mu$m).
Using a serial capacitor model we calculate a gate capacitance of 7.2~aF/$\mu$m$^{-2}$ which directly relates the carrier concentration to $V_G$ as $n = \alpha (V_G-V_{CNP})$ with leverage factor $\alpha = 0.5 \times 10^{14}$ m$^{-2}$V$^{-1}$ and a finite voltage of the the CNP of $V_{CNP}$~=~1.2~V.  In high magnetic fields, the geometric capacitance increases due to the formation of edge states\cite{Marun} and $\alpha$ becomes dependent on $B$. Therefore, the exact values of capacitance were determined experimentally by identifying the filling factors of quantized Hall plateaus in magnetic field, details can be found in the appendix.\\
After mounting the devices were slowly cooled down to 4.2~K and current annealed~\cite{Moser1} by applying a DC bias current up to 3~mA. This local annealing resulted into the high quality sample with mobility $\mu$ $\approx$ 10 m$^2$/Vs at a charge carrier density $n = 2\times 10^{11}~\text{cm}^{-2}$. The value of the mobility is calculated based on the dimension of suspended graphene before current annealing: 0.3~$\mu$m wide and 2.1~$\mu$m long. However, in the membrane the distribution of the temperature while current annealing is non homogenous,\cite{Tombros2} which most probably leads to the middle
part of the membrane being annealed and non annealed regions close by the contacts. In this case the estimation of the mobility value based on geometrical dimensions might be not precise. We can also estimate the quality of obtained sample from the value of magnetic field at which the system enters the quantum Hall regime (B~$>$~0.5~T). Assuming $\mu B \gg 1$ for QHE to exist\cite{Bolotin1}, the observation applies a lower bound for the mobility of 2~m$^2$/Vs.\\
Measurements were performed with standard low-frequency lock-in techniques in two-probe geometry with an excitation current of 2~nA.

\section{Results}

\begin{figure}[t]
\includegraphics[width=8cm]{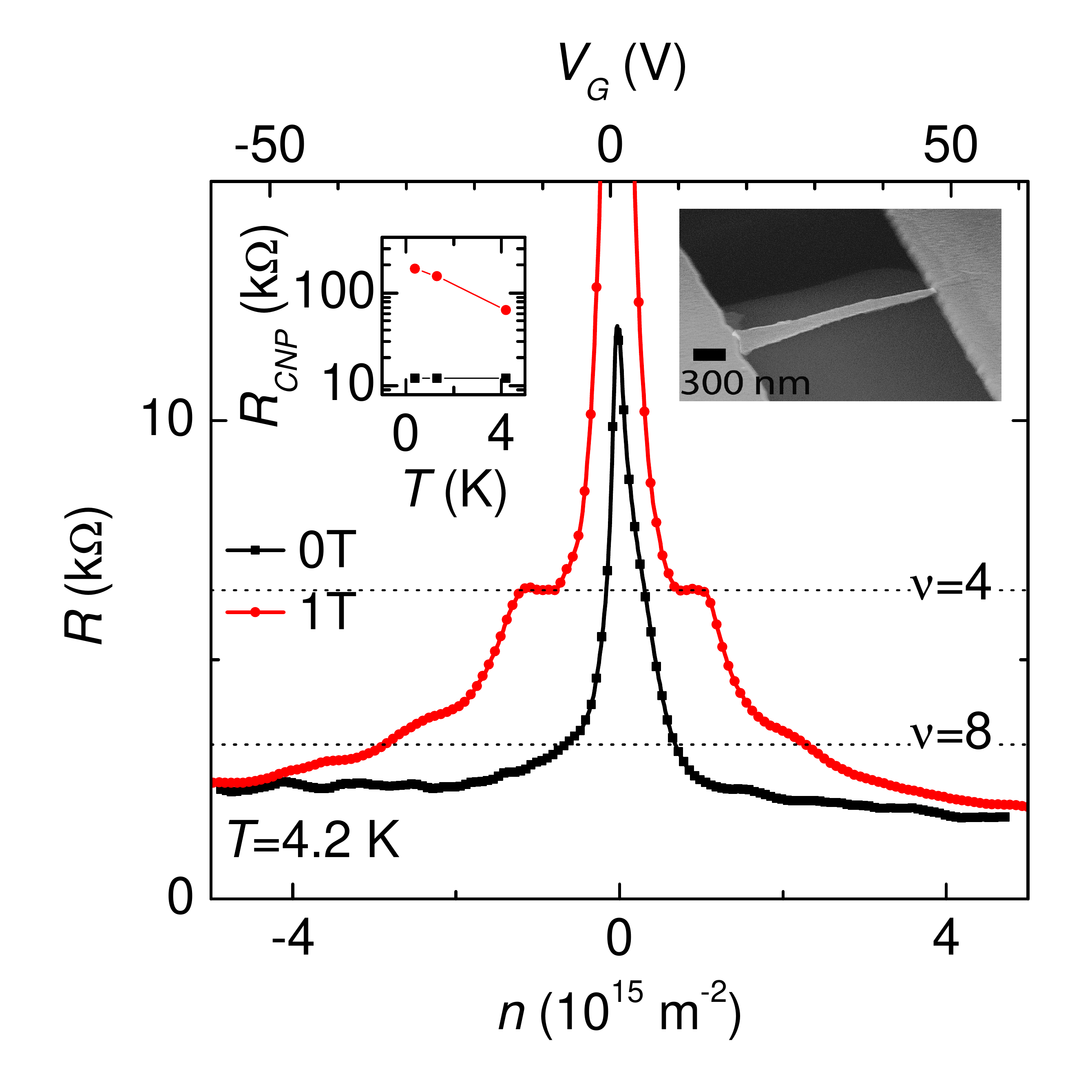}
\caption{Resistance as function of the concentration by sweeping the backgate from -60~V to 60~V for 0~T (black \small $\blacksquare$ \normalsize) and 1~T (red {\color{red}$\bullet$}). A constant contact resistance has been removed.  top-left inset: Temperature dependence of the resistance at the CNP for 0~T and 1~T; top-right inset: SEM-picture of our suspended device.}
\label{Fig1}
\end{figure}

In Fig.~\ref{Fig1} we show the data for the two-point resistance $R$ of our suspended BLG device at $B$~=~0~T and $B$~=~1~T as a function of $V_G$ (top x-axis) and $n$ (bottom x-axis), respectively. The two-probe resistance $R$ is characterized by a magnetoresistance $\rho_{xx} = L/w \cdot R_{xx}$ with superimposed Hall-resistance $\rho_{xy}$, \mbox{$R = (L/w) \cdot \rho_{xx} + \rho_{xy}$}. Here $L/w \approx 6.7$ is the aspect ratio of the device. The traces are corrected by phenomenological contact resistances (1~k$\Omega$ on the electron-side and 1.7~k$\Omega$ on hole-side) which were determined from a finite resistance background observed at high carrier concentrations; this background resistance increases by about a factor 2 in the range $B$~=~0...30~T. These contact resistances most probably originate from in-series connected non-annealed parts of the sample,\cite{Abanin1} contact doping\cite{Huard1,Blake2} and the finite resistance of the current leads. The sharp maximum at the CNP of the zero-field data already indicates the high electronic quality of the sample. At 1~T the resistance already exhibits fully quantized plateaus at filling factors $\nu=4$ and a developing quantization at $\nu=8$ and $\nu=12$. The formation of these plateaus is caused by a quantization of $\rho_{xy} = h/ \nu e^2$ and the associated zero minima in $\rho_{xx}$ when the Fermi energy lies between two Landau levels~\cite{Novoselov_UQHE} and confirms the high electronic mobility ($\mu \gg 1/B$) of our device required to observe this unconventional quantum Hall effect.

\begin{figure}[b!]
\includegraphics[width=8cm]{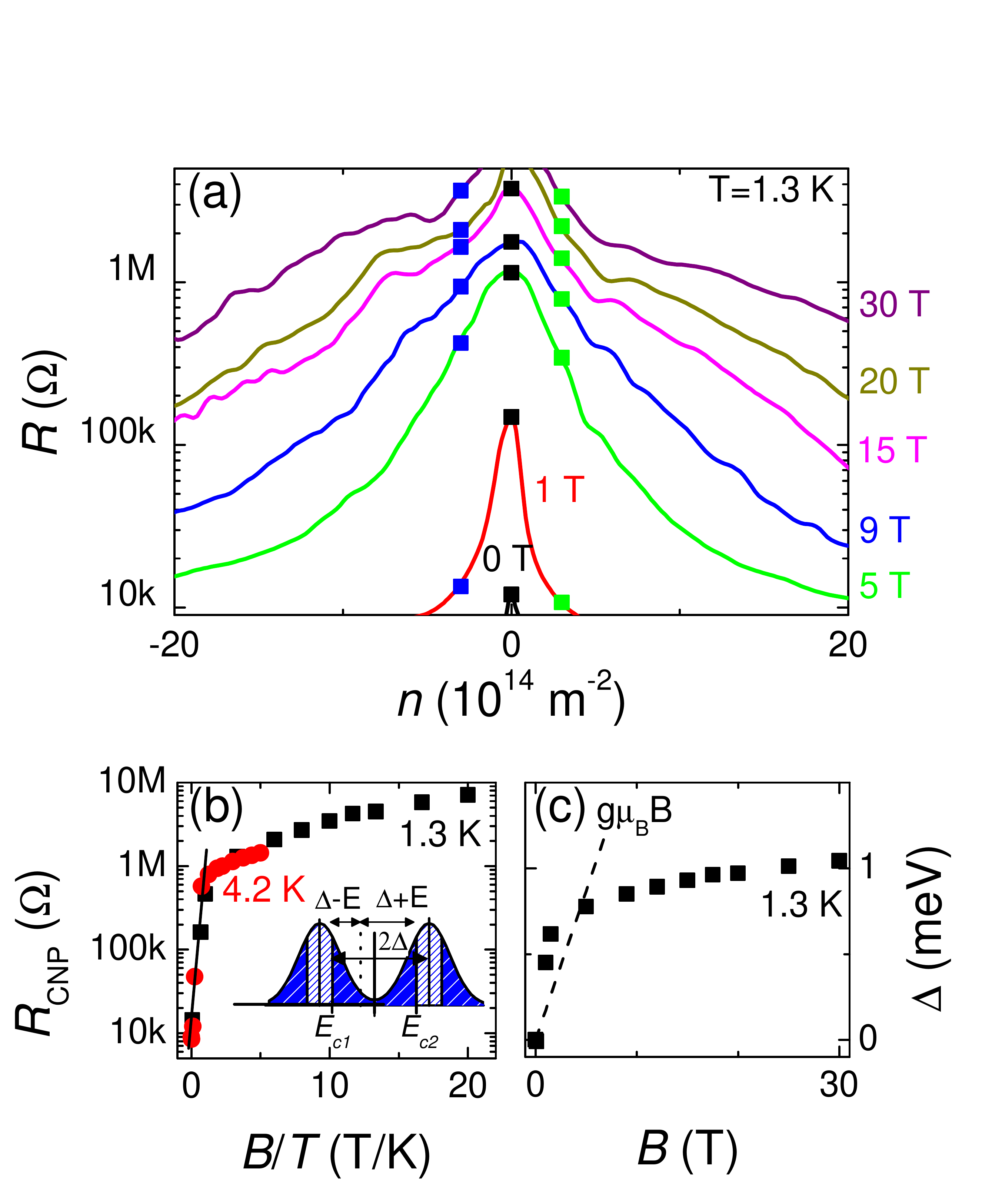}
\vspace*{3em}
\caption{(a) Resistive behaviour of the sample near the CNP at $T$=1.3~K. The dots mark the positions $n=3 \times 10^{14},0 \text{ and } -3 \times 10^{14}$ m$^{-2}$ where the gap opening has been analyzed (see text for more details). (b) Resistance $R_{CNP}$ as function of $B/T$ at $T$~=~4.2~K and 1.3~K; inset: Qualitative picture of the DOS near the CNP. The conduction at energy $E$ is directly related to the thermal excitation of electrons to the conduction edges $E_{c1}$ and $E_{c2}$. (c) Calculated gap $\Delta$ as function field $B$ for $T=1.3$~K; the dashed line represents the theoretical single electron Zeeman-energy $g\mu_B B$.}
\label{Fig2}
\end{figure}

Additionally, as soon as a finite magnetic field is applied, the resistance at the CNP, $R_{CNP}$, starts to diverge. Whereas at zero magnetic field $R_{CNP}$ is only very weakly temperature dependent and comparable to the resistance quantum, already at 1~T it is nearly an order of magnitude higher and starts to increase strongly with decreasing temperature, see left inset in Fig.~\ref{Fig1}. The nature of the gap opening at the CNP is elucidated further in Fig.~\ref{Fig2}a where we show the resistance as a function of carrier concentration $n$ for several magnetic fields. The diverging resistance at the CNP appears at similar magnetic field as the plateaus at filling factors -4 and 4; i.~e.~the eight-fold degeneracy of the zero-energy Landau level breaks directly into two four-fold degenerated Landau levels, as already predicted theoretically\cite{Mccannbilayers} and  proven experimentally.\cite{SeyoungKim1} At low fields $B<$~0.1~T we observe a small decrease of the resistance maximum at the CNP (not shown in the figure). This small decrease in resistance can be explained by the presence of local inhomogeneities which give a small splitting between the valley-polarized energies; the cross-over of these energy-states at finite magnetic field results in a resistance minimum. When the magnetic field is above $B\geq$~0.1~T we observe a rapid increase of the resistance-maximum at the CNP, shown in Fig.~\ref{Fig2}b. We can interpret this rapid increase as a result of the spin-splitting of the two energy levels at zero energy or by disorder, e.g. unevenly charged top and bottom layer. The last scenario would lead to a strong temperature dependence at zero field and ultimately for big disorder to an insulating state at zero field, as discussed in Ref. \onlinecite{Alina1}. The absence of a temperature influence at 0~T and the CNP centered at very low gate-voltage points to a non-disordered bilayer, therefore we interpret the rapid increase by a result of spin splitting.\\
The absence of an energy-level at $E=0$ in the inset of Fig.~\ref{Fig2}b results in a diverging resistance at the CNP. The resistance $R_{CNP}$ at the CNP follows a classical Ahrrenius-activation behaviour $R_{xx} \propto \exp{(\Delta/k_B T)}$, in where $\Delta$ is a scale for the size of the gap. The resistance increase scales best with $\ln{(R)} \propto B/T$, from which we obtain a gap $\Delta = \text{0.34~meV/T} \times B$. This gap is about a factor 3 times larger than the Zeeman-splitting $g\mu_B B$, which can be explained by the dominating exchange energy.\cite{Kurganova1} Equation (\ref{exchange-spin}) describes the total spin energy $\Delta_S$, determined by the sum of the single electron Zeeman energy $g\mu_B B$ and the exchange energy $E_{ex}\cdot (n_{\uparrow} - n_{\downarrow})$. Here $n_{\uparrow} - n_{\downarrow}$ is the normalized difference between spin-up and spin-down occupation.

\begin{equation}\label{exchange-spin}
\Delta_S=g\mu_B B + E_{ex} \cdot (n_{\uparrow} - n_{\downarrow})
\end{equation}

At low fields the two energy levels are still overlapping and the system is not fully spin polarized, $(n_{\uparrow} - n_{\downarrow}) < 1$. Assuming Gaussian shaped Landau levels we can approximate $(n_{\uparrow} - n_{\downarrow})=\frac{\sqrt{2}}{\pi}\frac{\Delta_S}{\Gamma}$ with leads with help of equation (\ref{exchange-spin}) to the gap  $\Delta_S=\frac{g \mu_B B}{1-E_{ex}/\Gamma}$. The observed spin-enhancement by a factor 3 corresponds to a typical level width $\Gamma$~=~2~meV and exchange energy $E_{ex}$~=~1.3~meV at $B$~=~1~T corresponding to a value of about 2~\% of the Coulomb energy $E_C=e^2/\epsilon_r l_B$~=~56~meV, where $l_B$ is the magnetic length.

The behavior at high magnetic field is experimentally more complicated to access, because the measured resistance rapidly exceeds several M$\Omega$s and a quantitative analysis becomes difficult. However away from the CNP the measured resistances stay low enough to guarantee a reliable interpretation up to the highest magnetic fields. This situation is illustrated in the inset of Fig.~\ref{Fig2}b where we sketch the quantized density of states in the lowest Landau level around the CNP with a gap 2$\Delta$ opening at $E=0$. When the Fermi energy is located at a finite energy $E < \Delta$ (i.e. still inside the localized parts of the DOS), conduction will occur by thermal excitation to the conductivity edges $E_{c1}$ and $E_{c2}$ of the extended states. The resistance $R(E)$ at this energy will then be given by
\begin{equation}\label{formula}
R(E)\propto e^{\frac{\Delta (E)+E}{k_bT}}+e^{\frac{\Delta (E)-E}{k_bT}}=\frac{1}{2}e^{\frac{\Delta (E)}{k_bT}}\cosh{\left(\frac{E}{k_bT}\right)}
\end{equation}

For relatively small energies $E<<k_bT$ the cosh-term can be approximated by a first order Taylor expansion $\cosh{\left(\frac{E}{k_B T}\right)}\approx 1 + \frac{1}{2}\frac{E^2}{k_B^2 T^2}=\gamma (E)$.  For small $E$ we can interpret equation (\ref{formula}) as $R_{CNP} \propto \frac{1}{2}e^{\frac{\Delta (E)}{k_bT}}\gamma (E)$. At the CNP, $E=0$, this approaches a trivial Ahrrenius-behavior, while for non-zero fixed energy $\gamma(E)$ is an energy dependent renormalization factor which for $E<<k_bT$ is independent of $T$.\\
We analyze the resistance at concentrations $n=\pm 3\times 10^{14} $~m$^{-2}$ (dots marked in Fig.~\ref{Fig2}a and multiply this data with a fixed constant to make an overlap with the low field data). All datapoints $R>~$1~M$\Omega$ in Fig.~\ref{Fig2}b are verified by this method and therefore reliable up to the highest field. This proper scaling for both low and high resistances also excludes a strong effect of the local heating due the finite excitation voltage we apply over the sample.\\
From Fig.~\ref{Fig2}b we see that the scaling of the resistance at high magnetic fields is remarkably different from the linear field increase at low fields. This observation is again visualized in Fig. \ref{Fig2}c where we show that the calculated gap strongly bends and the slope strongly reduces. In this regime the gatesweeps are packed more densly for increasing magnetic field and the energy $E$ gets  comparable to the thermal activation $k_bT$ thus we are no longer able to calculate the gap-size with a simple Arrhenius-behaviour. Experimental limitations of our suspended samples do not allow us to access much higher temperatures, therefore we can only speculate here about further gap-study.\\
At high enough fields we expect to fulfill the criteria of fully spin-polarized system, $(n_{\uparrow} - n_{\downarrow}) \rightarrow 1$. The sudden strong change of the gap-size suggests that our system indeed gets fully spin-polarized, in literature also known as the cross-over to a quantum Hall ferromagnetic state. Further increase of the magnetic field leads hypothetically to a dominating spin-splitting $g \mu_B B$, because $E_{ex} \propto \sqrt{B}$. Further specific research in titled magnetic fields is necessary to decouple the influence of the single electron Zeeman energy and exchange energy.\\

\begin{figure}[t]
\includegraphics[width=8cm]{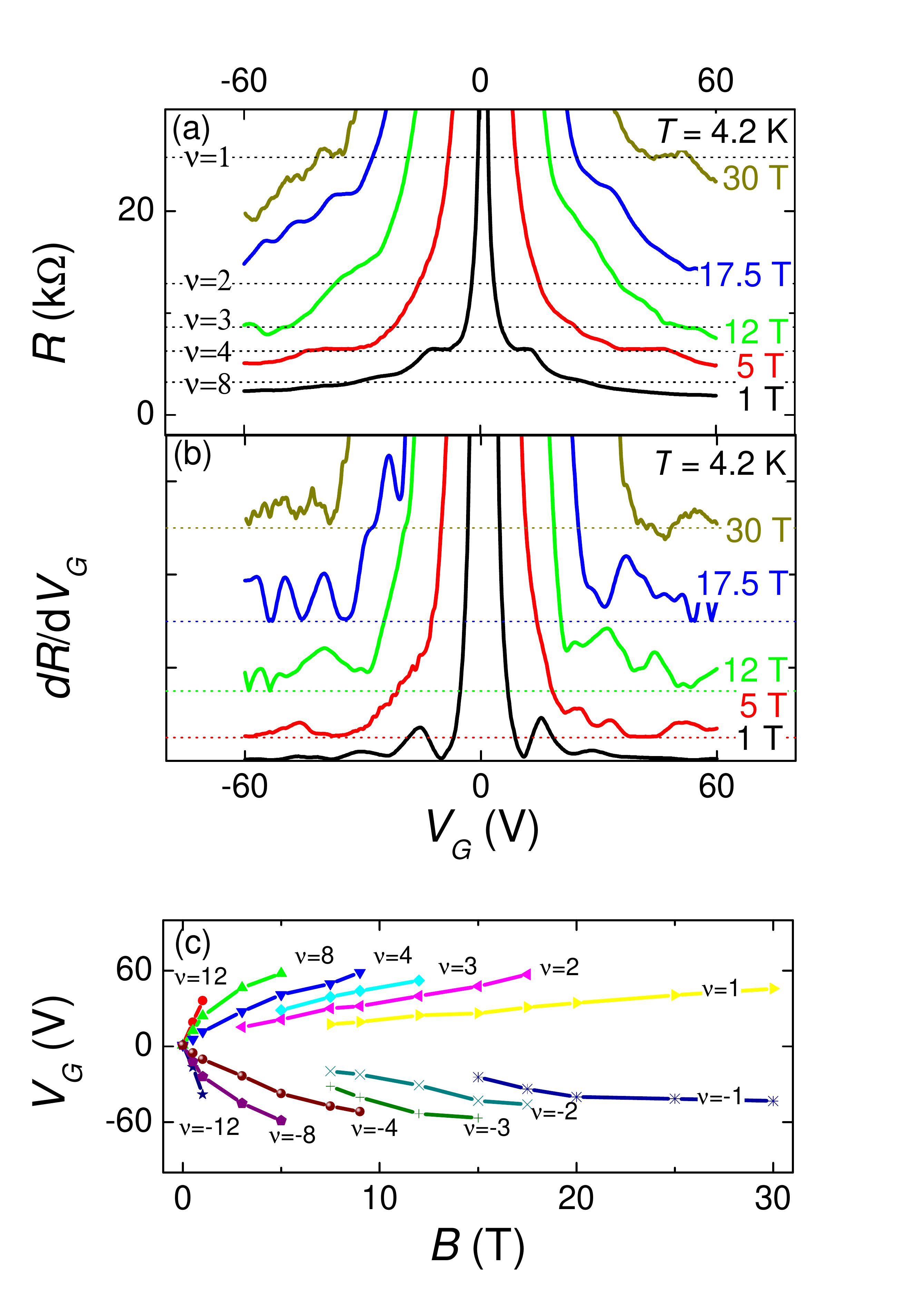}
\caption{(a) Gatesweeps $R(V_G)$ at constant magnetic fields $B$~=~1~T, 5~T, 12~T, 17.5~T and 30~T for $T$~=~4.2~K; the curves are shifted up for clarity. (b) Derivative d$R$/d$V_G$ for the curves in (a). (c) Position of the minima for $\nu$~=~$\pm$12, $\pm$8, $\pm$4, $\pm$3, $\pm$2 and $\pm$1 as function of the magnetic field.}
\label{Fig3}
\end{figure}

After detailed study at low concentrations we have a closer look at the manifestation of the QHE at higher concentration. In Fig.~\ref{Fig3}a we show the corrected two-point resistance as a function $V_G$ at 4.2~K for $B$~=~1~T, 5~T, 12~T, 17.5~T and 30~T. Apart from the distinct $\nu= \pm 4$ plateaus which are already well pronounced at 1~T, additional plateaus at $\nu= \pm 3, \pm 2,$ and $\pm 1$ start to appear in higher fields. In Fig.~\ref{Fig3}b we show the derivative $\left|\frac{\text{d}R}{\text{d}V_G}\right|$ of the resistance curves, where we can already already recognize distinct maxima and minima for lower fields.

In Fig.~\ref{Fig3}c we follow the position of the minima with increasing magnetic field. We see that the maximum applied gate-voltage $V_G$~=~60~V limits the observation of filling factors $\nu = \pm 4$ up to 9~T, while $\nu = \pm 2$ remains observable till fields of 15~-~20~T and filling factor $\nu = \pm 1$ is still observable at the highest applied field, 30~T. From Fig.~\ref{Fig3}c we observe that the position of the minima strongly deviates from the linear relationship between the induced charge carrier concentration $n$ and the applied magnetic field $B$, $n=\nu\frac{eB}{h}$. The equidistance of the minima for fixed magnetic fields excludes a capacitance-change due the bending of the membrane; which can be expected by the particular big difference between the electric field induced bending (10~-~20~nm)\cite{Wang1} and the vacuum gap over which graphene is suspended ($\sim 1.5~\mu$m). In the appendix we discuss in more detail how to extract the exact relation between the applied field $B$ and the induced charge concentration $n$ from this data.
\begin{figure}[t]
\includegraphics[width=8cm]{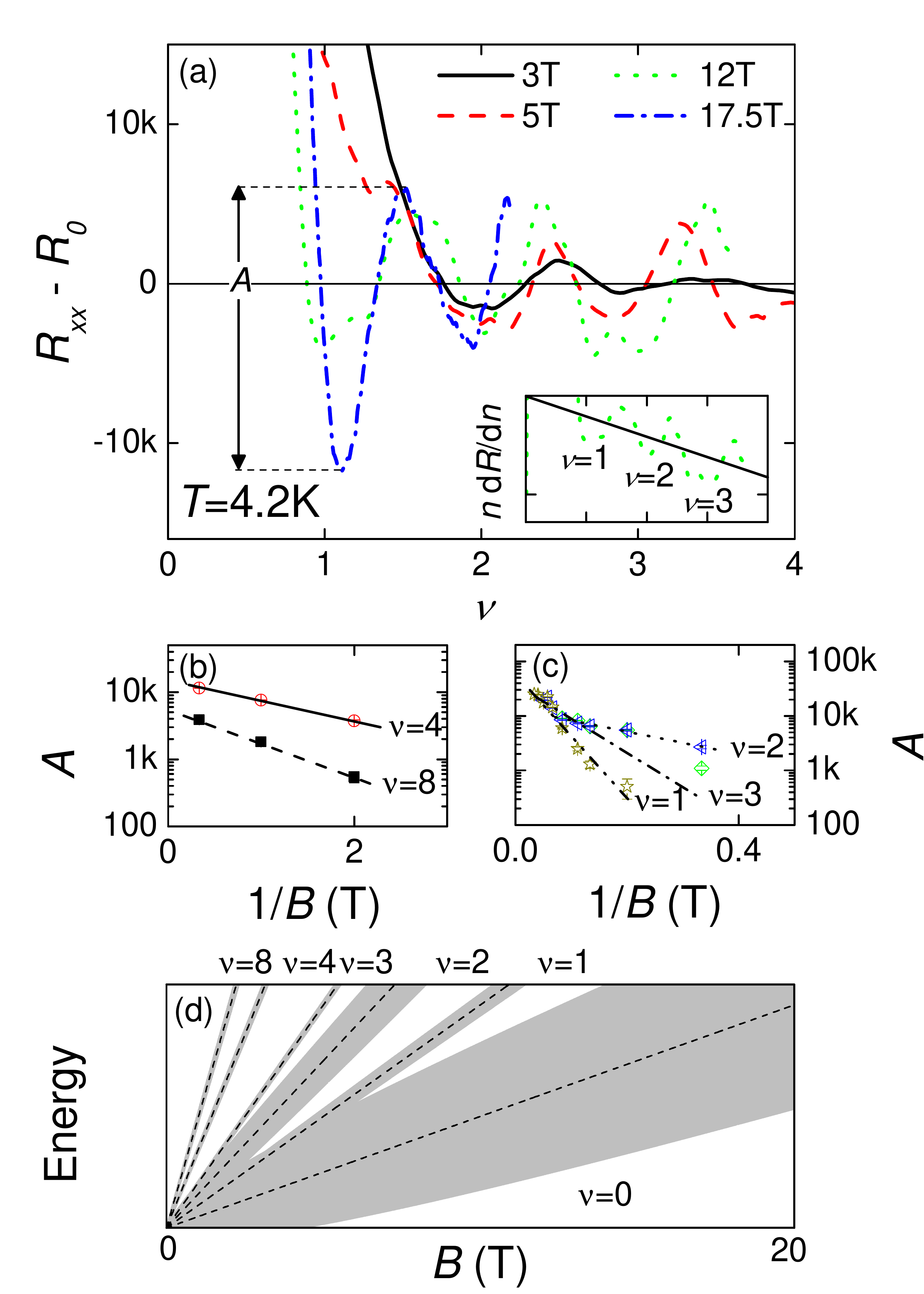}
\caption{(a) $R_{xx}$-oscillations after removing linear background from $n\frac{\text{d}R}{\text{d}n}$ for $B$~=~3~T, 5~T, 12~T and 20~T at 4.2~K. (b) Dingle plot of $\nu$~=~8 and $\nu$~=~4: amplitude $A$ of the oscillations as function of the inverse field $1/B$. (c) Dingle plot for $\nu$~=~3,~2 and 1. (d) Schematic plot of the appearance filling factors with increasing magnetic field. The dashed lines show the center of the Landau level, while the grey shaded area is the Landau level broadening determined by the Dingle temperature $T_D$.}
\label{Fig4}
\end{figure}

Experiments on suspended graphene-samples are mainly performed in two-probe configuration. Experimental limitations of the annealing-procedure do not allow us to obtain very homogenous samples in four-probe configuration. Therefore more effort has to be done to do a proper analysis on both the magnetoresistance and Hall-resistance. In Fig.~\ref{Fig4} the appearance of the different filling factors are further elucidated; in particular at positive gate-voltages, the influence of contact-resistance is here experimentally the smallest. As shown in Fig.~\ref{Fig3}b the derivative of our data, $\frac{\text{d}R}{\text{d}V_G}$, shows already at very low field a very clear appearance of filling factors $\nu$~=~1, 2, 3 and 4. A small change in the slope of $R_{xy}$ causes a very distinct minimum in the derivative. Theoretically we can use a model that directly describes the magnetoresistance $R_{xx}$ in terms of the Hall resistance $R_{xy}$,\cite{Stormer1} i.e. $R_{xx} \propto n\frac{\text{d}R_{xy}}{\text{d}n}$. In Fig.~\ref{Fig4}a we study the appearance of the filling factors by plotting the obtained magnetoresistances $R_{xx}-R_0$. Here we removed from all data the linear background $R_0$ of the 12~T data shown in the inset of Fig.~\ref{Fig4}a and centered all curves around the x-axis. We used the obtained leverage factor $\alpha(B)$ to determine the exact concentration $n$. Already at 3~T we observe the appearance of clear oscillations around $\nu=2$ and $\nu=3$ followed by the appearance of $\nu=1$ at 5~T. The amplitude $A$ of the oscillation is defined by the difference between the minimum and the first maximum. A single oscillation can be best analyzed by applying Lifshitz-Kosevic equation\cite{Kosevic} $A\cdot \cos{(f(B))}$, here $A$ is the amplitude and $f(B)$ a field-dependent function that determines the frequency and phase. The amplitude $A$ is finite due to the Landau-level broadening, and is damped by the Dingle-factor $\exp{(-\beta \cdot T_D m_c/B)}$ in where $T_D$ is the Dingle-temperature, $m_c$ the cyclotron mass in units of electron mass $m_e$, $B$ the magnetic field and $\beta=14.694$~T/K. In Fig.~\ref{Fig4}b the amplitudes $A$ for $\nu=4$ and $\nu=8$ are plotted as function of $1/B$, which affects in a linear decrease with slope $\beta \cdot T_D m_c$. If we assume the cyclotron mass in bilayer graphene to be $m_c \approx 0.033 \cdot m_e$\cite{} (corresponding to $\gamma_1$~=~0.4~eV, see Ref.~\onlinecite{Kurganova2} and references in there) we obtain the Dingle temperatures $T_D$ in the table below. We repeat the same procedure for filling factors 1, 2 and 3 in Fig.~\ref{Fig4}c.\\
\begin{center}
\begin{tabular}{|c|c|c|c|c|c|}
\hline
$\nu$ & 8 & 4 & 3 & 2 & 1\\
\hline
$T_D$~(K) & $2.4 \pm 0.4$ & $1.4 \pm 0.4$ & $29.2 \pm 4$ & $9.2\pm 1.2$ & $58 \pm 8$\\
\hline
\end{tabular}
\end{center}
Compared to $\nu=4$, fully quantized at $B=1$~T, the Dingle temperatures $T_D$ for the degenerate filling factors are one order of magnitude larger, which means fields $B\geq$~10~T are required to observe full quantization; in particular filling factor $\nu=1$ becomes quantized at fields of $B \gtrapprox$~30~T.\\
In Fig.~\ref{Fig4} we illustrate qualitatively the appearance of the different Landau-levels for increasing magnetic field. The corresponding grey shaded areas describe the Landau level broadening $\Gamma$, directly determined by the Dingle temperature $T_D$; higher Dingle temperatures correspond broader Landau levels. While the position of the energy moves linearly with increasing field, the Landau level broadening $\Gamma$ is proportional to the square root of the applied field $\Gamma \propto \sqrt{B}$. With increasing field the overlap between the shaded areas decreases, and the plateau starts to appear. As we can see from Fig.~\ref{Fig4}d Landau levels around $\nu=2$ and $\nu=3$ do indeed not overlap anymore for similar magnetic field, however the shaded areas for $\nu=1$ overlap till higher fields. Finally the overlapping of filling factors $\nu=\pm 1$ disappears at similar magnetic field as the resistance at the CNP starts to bend strongly, supporting the idea of a cross-over to a fully spin polarized state at $\nu=0$.\\
After the appearance of the non-degenerated filling factors $\nu=4$ and 8 a gap at $\nu=0$ forms, followed by filling factors $\nu=2$ and at high fields $\nu=1$ and 3. This sequence agrees with the proposed model of the formation of a quantum Hall ferromagnet in the lowest Landau level and the observed behaviour at the CNP.

\section{Conclusion}
In conclusion we have performed experiments on a suspended BLG sample which shows us a field induced gap at the CNP for fields $B>1$~T. The gap at $\nu=0$ opens simultaneously with the formation of filing factors $\nu=\pm 4$, which implicates the eight-fold degenerated lowest Landau breaks directly in two fourfold-degenerated spin-polarized subbands. At high magnetic fields we observe a smooth transition to a much smaller gap, this is consistent with the picture of the formation of a spin-polarized quantum Hall ferromagnetic state. Following the breaking of the lowest Landau level we observe a breaking of $\nu=0$ in $\nu=\pm2$ and finally in $\nu=\pm1$ and $\nu=\pm3$, in agreement with the theoretically proposed model of a quantum Hall ferromagnet.\\

\begin{acknowledgments}
This work is part of the research program of the 'Stichting voor Fundamenteel Onderzoek der Materie (FOM)', which is financially supported by the 'Nederlandse Organisatie voor Wetenschappelijk Onderzoek (NWO)'. We also thank Zernike for Advanced Materials Institute and Nanoned for financial support.\\
\\
\end{acknowledgments}

\section{Appendix}
\begin{figure}[h!]
\includegraphics[width=8cm]{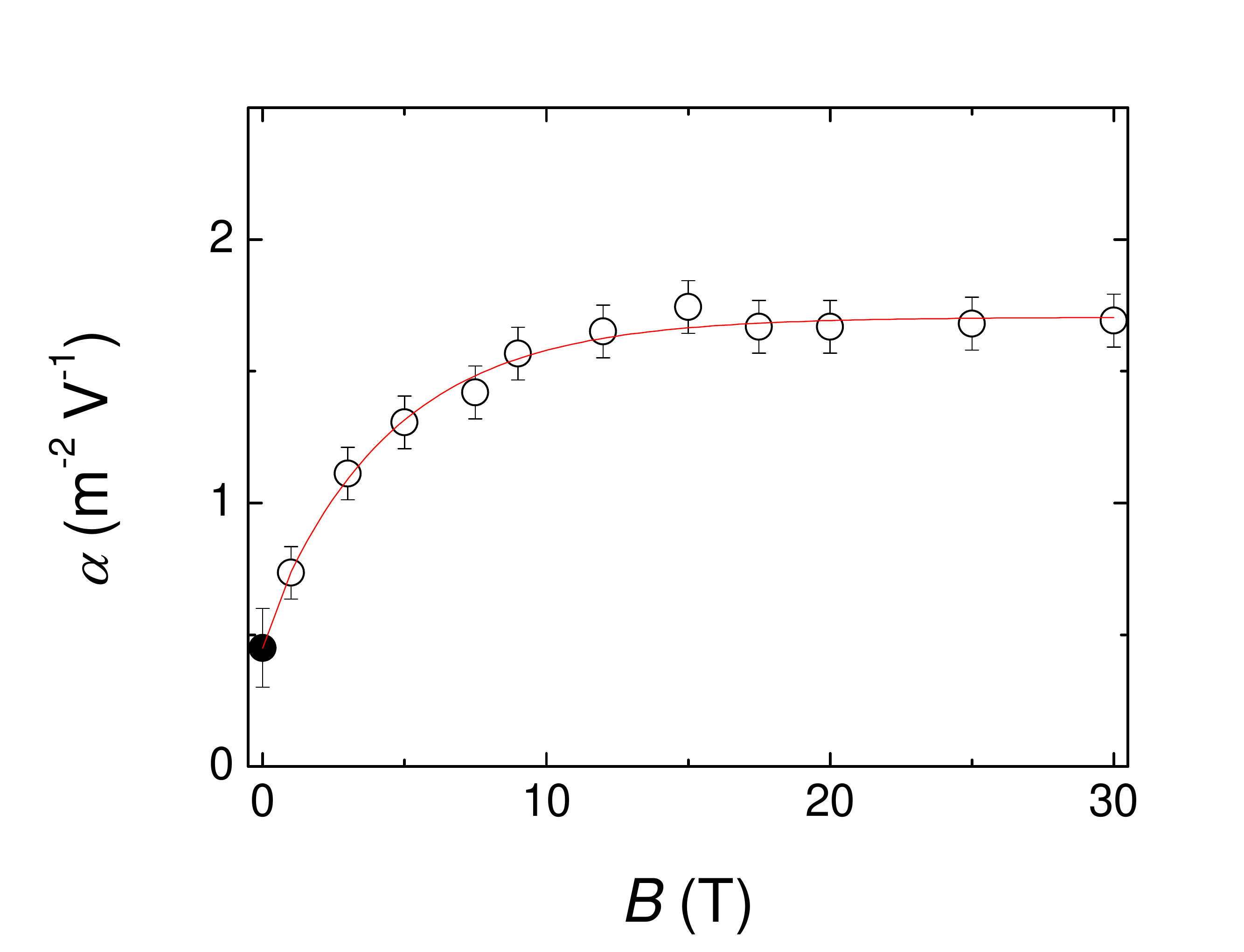}
\caption{Relation between the induced charge carrier concentration per applied magnetic field $\alpha(B)=n/B$ and the applied magnetic field $B$ for $B$=0 ($\bullet$) and $B\neq$0 ($\circ$). The plotted line shows the interpolated $\alpha (B)$ from which we determined a reliable value of the concentration $n$.}
\label{Fig1supp}
\end{figure}

Fig.~\ref{Fig1supp} shows the relation between $\alpha$ and the applied magnetic field. The capacitance of the sample increases from the geometrical value 0.5~$\times 10^{14}$~m$^{-2}$~V~$^{-1}$ up to 1.8~$\times 10^{14}$~m$^{-2}$~V~$^{-1}$ at 9~T and saturates at this value for the highest fields. This effect is also observed implicitly in recent publications \cite{Feldman1,Freitag} on high quality suspended bilayer-devices, but not mentioned by authors in the text. \\
As discussed in ref. \onlinecite{Marun} the increase in capacitance of the system under applied magnetic field could be understood from the deviation from the flat-plate capacitor model. At the point when the width of the graphene flake is smaller or comparable to the distance to the back gate the flat-plate capacitor model can be no longer applied. The charge carrier distribution in graphene becomes non homogenous and increases at the edges. Since the classical cyclotron radius of charge carrier depends inversely on magnetic field, the increase of $B$ will cause edge channels in quantum Hall regime propagate closer to the edge, where the density can be few times higher than in bulk graphene. This would result an increase in capacitance extracted from QHE plateaus. The cyclotron radius is expected to be dependent on the charge carrier density as well. The exact calculations for different device's geometries with charge carrier distribution in graphene, compared to the experiments, are the subject of another paper.\cite{Marun}


\begin{thebibliography}{34}%
\makeatletter
\providecommand \@ifxundefined [1]{%
 \@ifx{#1\undefined}
}%
\providecommand \@ifnum [1]{%
 \ifnum #1\expandafter \@firstoftwo
 \else \expandafter \@secondoftwo
 \fi
}%
\providecommand \@ifx [1]{%
 \ifx #1\expandafter \@firstoftwo
 \else \expandafter \@secondoftwo
 \fi
}%
\providecommand \natexlab [1]{#1}%
\providecommand \enquote  [1]{``#1''}%
\providecommand \bibnamefont  [1]{#1}%
\providecommand \bibfnamefont [1]{#1}%
\providecommand \citenamefont [1]{#1}%
\providecommand \href@noop [0]{\@secondoftwo}%
\providecommand \href [0]{\begingroup \@sanitize@url \@href}%
\providecommand \@href[1]{\@@startlink{#1}\@@href}%
\providecommand \@@href[1]{\endgroup#1\@@endlink}%
\providecommand \@sanitize@url [0]{\catcode `\\12\catcode `\$12\catcode
  `\&12\catcode `\#12\catcode `\^12\catcode `\_12\catcode `\%12\relax}%
\providecommand \@@startlink[1]{}%
\providecommand \@@endlink[0]{}%
\providecommand \url  [0]{\begingroup\@sanitize@url \@url }%
\providecommand \@url [1]{\endgroup\@href {#1}{\urlprefix }}%
\providecommand \urlprefix  [0]{URL }%
\providecommand \Eprint [0]{\href }%
\@ifxundefined \urlstyle {%
  \providecommand \doi  [0]{\begingroup \@sanitize@url \@doi}%
  \providecommand \@doi [1]{\endgroup \@@startlink {\doibase
  #1}doi:\discretionary {}{}{}#1\@@endlink }%
}{%
  \providecommand \doi  [0]{doi:\discretionary{}{}{}\begingroup
  \urlstyle{rm}\Url }%
}%
\providecommand \doibase [0]{http://dx.doi.org/}%
\providecommand \Doi [0]{\begingroup \@sanitize@url \@Doi }%
\providecommand \@Doi  [1]{\endgroup\@@startlink{\doibase#1}\@@Doi}%
\providecommand \@@Doi [1]{#1\@@endlink}%
\providecommand \selectlanguage [0]{\@gobble}%
\providecommand \bibinfo  [0]{\@secondoftwo}%
\providecommand \bibfield  [0]{\@secondoftwo}%
\providecommand \translation [1]{[#1]}%
\providecommand \BibitemOpen [0]{}%
\providecommand \bibitemStop [0]{}%
\providecommand \bibitemNoStop [0]{.\EOS\space}%
\providecommand \EOS [0]{\spacefactor3000\relax}%
\providecommand \BibitemShut  [1]{\csname bibitem#1\endcsname}%
\bibitem [{\citenamefont {Geim}\ and\ \citenamefont {Novoselov}(2007)}]{Geim1}%
  \BibitemOpen
  \bibfield  {author} {\bibinfo {author} {\bibfnamefont {A.~K.}\ \bibnamefont
  {Geim}}\ and\ \bibinfo {author} {\bibfnamefont {K.~S.}\ \bibnamefont
  {Novoselov}},\ }\href@noop {} {\bibfield  {journal} {\bibinfo  {journal}
  {Nature Materials},\ }\textbf {\bibinfo {volume} {6}},\ \bibinfo {pages}
  {183} (\bibinfo {year} {2007})}\BibitemShut {NoStop}%
\bibitem [{\citenamefont {Bolotin}\ \emph {et~al.}(2008)\citenamefont
  {Bolotin}, \citenamefont {Sikes}, \citenamefont {Jiang}, \citenamefont
  {Klima}, \citenamefont {Fudenberg}, \citenamefont {Hone}, \citenamefont
  {Kim},\ and\ \citenamefont {Stormer}}]{Bolotin1}%
  \BibitemOpen
  \bibfield  {author} {\bibinfo {author} {\bibfnamefont {K.~I.}\ \bibnamefont
  {Bolotin}}, \bibinfo {author} {\bibfnamefont {K.~J.}\ \bibnamefont {Sikes}},
  \bibinfo {author} {\bibfnamefont {Z.}~\bibnamefont {Jiang}}, \bibinfo
  {author} {\bibfnamefont {M.}~\bibnamefont {Klima}}, \bibinfo {author}
  {\bibfnamefont {G.}~\bibnamefont {Fudenberg}}, \bibinfo {author}
  {\bibfnamefont {J.}~\bibnamefont {Hone}}, \bibinfo {author} {\bibfnamefont
  {P.}~\bibnamefont {Kim}}, \ and\ \bibinfo {author} {\bibfnamefont {H.~L.}\
  \bibnamefont {Stormer}},\ }\href@noop {} {\bibfield  {journal} {\bibinfo
  {journal} {Solid State Communications},\ }\textbf {\bibinfo {volume} {146}},\
  \bibinfo {pages} {351} (\bibinfo {year} {2008})}\BibitemShut {NoStop}%
\bibitem [{\citenamefont {Du}\ \emph {et~al.}(2008)\citenamefont {Du},
  \citenamefont {Skachko},\ and\ \citenamefont {Andrei}}]{Du1}%
  \BibitemOpen
  \bibfield  {author} {\bibinfo {author} {\bibfnamefont {X.}~\bibnamefont
  {Du}}, \bibinfo {author} {\bibfnamefont {I.}~\bibnamefont {Skachko}}, \ and\
  \bibinfo {author} {\bibfnamefont {E.~Y.}\ \bibnamefont {Andrei}},\
  }\href@noop {} {\bibfield  {journal} {\bibinfo  {journal} {International
  Journal of Modern Physics B},\ }\textbf {\bibinfo {volume} {22}},\ \bibinfo
  {pages} {4579} (\bibinfo {year} {2008})}\BibitemShut {NoStop}%
\bibitem [{\citenamefont {Dean}\ \emph {et~al.}(2010)\citenamefont {Dean},
  \citenamefont {Young}, \citenamefont {Meric}, \citenamefont {Lee},
  \citenamefont {Wang}, \citenamefont {Sorgenfrei}, \citenamefont {Watanabe},
  \citenamefont {Taniguchi}, \citenamefont {Kim}, \citenamefont {Shepard},\
  and\ \citenamefont {Hone}}]{Dean1}%
  \BibitemOpen
  \bibfield  {author} {\bibinfo {author} {\bibfnamefont {C.~R.}\ \bibnamefont
  {Dean}}, \bibinfo {author} {\bibfnamefont {A.~F.}\ \bibnamefont {Young}},
  \bibinfo {author} {\bibfnamefont {I.}~\bibnamefont {Meric}}, \bibinfo
  {author} {\bibfnamefont {C.}~\bibnamefont {Lee}}, \bibinfo {author}
  {\bibfnamefont {L.}~\bibnamefont {Wang}}, \bibinfo {author} {\bibfnamefont
  {S.}~\bibnamefont {Sorgenfrei}}, \bibinfo {author} {\bibfnamefont
  {K.}~\bibnamefont {Watanabe}}, \bibinfo {author} {\bibfnamefont
  {T.}~\bibnamefont {Taniguchi}}, \bibinfo {author} {\bibfnamefont
  {P.}~\bibnamefont {Kim}}, \bibinfo {author} {\bibfnamefont {K.~L.}\
  \bibnamefont {Shepard}}, \ and\ \bibinfo {author} {\bibfnamefont
  {J.}~\bibnamefont {Hone}},\ }\href@noop {} {\bibfield  {journal} {\bibinfo
  {journal} {Nature Nanotechnology},\ }\textbf {\bibinfo {volume} {5}},\
  \bibinfo {pages} {722} (\bibinfo {year} {2010})}\BibitemShut {NoStop}%
\bibitem [{\citenamefont {Dean}\ \emph {et~al.}(2011)\citenamefont {Dean},
  \citenamefont {Young}, \citenamefont {Cadden-Zimansky}, \citenamefont {Wang},
  \citenamefont {Ren}, \citenamefont {Watanabe}, \citenamefont {Taniguchi},
  \citenamefont {Kim}, \citenamefont {Hone},\ and\ \citenamefont
  {Shepard}}]{Dean2}%
  \BibitemOpen
  \bibfield  {author} {\bibinfo {author} {\bibfnamefont {C.~R.}\ \bibnamefont
  {Dean}}, \bibinfo {author} {\bibfnamefont {A.~F.}\ \bibnamefont {Young}},
  \bibinfo {author} {\bibfnamefont {P.}~\bibnamefont {Cadden-Zimansky}},
  \bibinfo {author} {\bibfnamefont {L.}~\bibnamefont {Wang}}, \bibinfo {author}
  {\bibfnamefont {H.}~\bibnamefont {Ren}}, \bibinfo {author} {\bibfnamefont
  {K.}~\bibnamefont {Watanabe}}, \bibinfo {author} {\bibfnamefont
  {T.}~\bibnamefont {Taniguchi}}, \bibinfo {author} {\bibfnamefont
  {P.}~\bibnamefont {Kim}}, \bibinfo {author} {\bibfnamefont {J.}~\bibnamefont
  {Hone}}, \ and\ \bibinfo {author} {\bibfnamefont {K.~L.}\ \bibnamefont
  {Shepard}},\ }\href@noop {} {\bibfield  {journal} {\bibinfo  {journal}
  {Nature Physics},\ }\textbf {\bibinfo {volume} {7}},\ \bibinfo {pages}
  {693–696} (\bibinfo {year} {2011})}\BibitemShut {NoStop}%
\bibitem [{\citenamefont {Bolotin}\ \emph {et~al.}(2011)\citenamefont
  {Bolotin}, \citenamefont {Ghahari}, \citenamefont {Shulman}, \citenamefont
  {Stormer},\ and\ \citenamefont {Kim}}]{Bolotin2}%
  \BibitemOpen
  \bibfield  {author} {\bibinfo {author} {\bibfnamefont {K.~I.}\ \bibnamefont
  {Bolotin}}, \bibinfo {author} {\bibfnamefont {F.}~\bibnamefont {Ghahari}},
  \bibinfo {author} {\bibfnamefont {M.~D.}\ \bibnamefont {Shulman}}, \bibinfo
  {author} {\bibfnamefont {H.~L.}\ \bibnamefont {Stormer}}, \ and\ \bibinfo
  {author} {\bibfnamefont {P.}~\bibnamefont {Kim}},\ }\href@noop {} {\bibfield
  {journal} {\bibinfo  {journal} {Nature},\ }\textbf {\bibinfo {volume} {475}}
  (\bibinfo {year} {2011})}\BibitemShut {NoStop}%
\bibitem [{\citenamefont {Du}\ \emph {et~al.}(2009)\citenamefont {Du},
  \citenamefont {Skachko}, \citenamefont {Duerr}, \citenamefont {Luican},\ and\
  \citenamefont {Andrei}}]{Du2}%
  \BibitemOpen
  \bibfield  {author} {\bibinfo {author} {\bibfnamefont {X.}~\bibnamefont
  {Du}}, \bibinfo {author} {\bibfnamefont {I.}~\bibnamefont {Skachko}},
  \bibinfo {author} {\bibfnamefont {F.}~\bibnamefont {Duerr}}, \bibinfo
  {author} {\bibfnamefont {A.}~\bibnamefont {Luican}}, \ and\ \bibinfo {author}
  {\bibfnamefont {E.~Y.}\ \bibnamefont {Andrei}},\ }\href@noop {} {\bibfield
  {journal} {\bibinfo  {journal} {Nature},\ }\textbf {\bibinfo {volume}
  {462}},\ \bibinfo {pages} {192} (\bibinfo {year} {2009})}\BibitemShut
  {NoStop}%
\bibitem [{\citenamefont {Feldman}\ \emph {et~al.}(2009)\citenamefont
  {Feldman}, \citenamefont {Martin},\ and\ \citenamefont {Yacoby}}]{Feldman1}%
  \BibitemOpen
  \bibfield  {author} {\bibinfo {author} {\bibfnamefont {B.~E.}\ \bibnamefont
  {Feldman}}, \bibinfo {author} {\bibfnamefont {J.}~\bibnamefont {Martin}}, \
  and\ \bibinfo {author} {\bibfnamefont {A.}~\bibnamefont {Yacoby}},\
  }\href@noop {} {\bibfield  {journal} {\bibinfo  {journal} {Nature Physics},\
  }\textbf {\bibinfo {volume} {5}},\ \bibinfo {pages} {889} (\bibinfo {year}
  {2009})}\BibitemShut {NoStop}%
\bibitem [{\citenamefont {Tombros}\ \emph
  {et~al.}(2011){\natexlab{a}}\citenamefont {Tombros}, \citenamefont
  {Veligura}, \citenamefont {Junesch}, \citenamefont {Guimaraes}, \citenamefont
  {Vera-Marun}, \citenamefont {Jonkman},\ and\ \citenamefont {van
  Wees}}]{Tombros2}%
  \BibitemOpen
  \bibfield  {author} {\bibinfo {author} {\bibfnamefont {N.}~\bibnamefont
  {Tombros}}, \bibinfo {author} {\bibfnamefont {A.}~\bibnamefont {Veligura}},
  \bibinfo {author} {\bibfnamefont {J.}~\bibnamefont {Junesch}}, \bibinfo
  {author} {\bibfnamefont {M.~H.~D.}\ \bibnamefont {Guimaraes}}, \bibinfo
  {author} {\bibfnamefont {I.~J.}\ \bibnamefont {Vera-Marun}}, \bibinfo
  {author} {\bibfnamefont {H.~T.}\ \bibnamefont {Jonkman}}, \ and\ \bibinfo
  {author} {\bibfnamefont {B.~J.}\ \bibnamefont {van Wees}},\ }\href@noop {}
  {\bibfield  {journal} {\bibinfo  {journal} {Nature Physics},\ }\textbf
  {\bibinfo {volume} {7}},\ \bibinfo {pages} {697} (\bibinfo {year}
  {2011}{\natexlab{a}})}\BibitemShut {NoStop}%
\bibitem [{\citenamefont {Novoselov}\ \emph {et~al.}(2006)\citenamefont
  {Novoselov}, \citenamefont {McCann}, \citenamefont {Morozov}, \citenamefont
  {Fal/'ko}, \citenamefont {Katsnelson}, \citenamefont {Zeitler}, \citenamefont
  {Jiang}, \citenamefont {Schedin},\ and\ \citenamefont
  {Geim}}]{Novoselov_UQHE}%
  \BibitemOpen
  \bibfield  {author} {\bibinfo {author} {\bibfnamefont {K.~S.}\ \bibnamefont
  {Novoselov}}, \bibinfo {author} {\bibfnamefont {E.}~\bibnamefont {McCann}},
  \bibinfo {author} {\bibfnamefont {S.~V.}\ \bibnamefont {Morozov}}, \bibinfo
  {author} {\bibfnamefont {V.~I.}\ \bibnamefont {Fal/'ko}}, \bibinfo {author}
  {\bibfnamefont {M.~I.}\ \bibnamefont {Katsnelson}}, \bibinfo {author}
  {\bibfnamefont {U.}~\bibnamefont {Zeitler}}, \bibinfo {author} {\bibfnamefont
  {D.}~\bibnamefont {Jiang}}, \bibinfo {author} {\bibfnamefont
  {F.}~\bibnamefont {Schedin}}, \ and\ \bibinfo {author} {\bibfnamefont
  {A.~K.}\ \bibnamefont {Geim}},\ }\href@noop {} {\bibfield  {journal}
  {\bibinfo  {journal} {Nature Physics},\ }\textbf {\bibinfo {volume} {2}},\
  \bibinfo {pages} {177} (\bibinfo {year} {2006})}\BibitemShut {NoStop}%
\bibitem [{\citenamefont {Zhao}\ \emph {et~al.}(2010)\citenamefont {Zhao},
  \citenamefont {Cadden-Zimansky}, \citenamefont {Jiang},\ and\ \citenamefont
  {Kim}}]{Zhao1}%
  \BibitemOpen
  \bibfield  {author} {\bibinfo {author} {\bibfnamefont {Y.}~\bibnamefont
  {Zhao}}, \bibinfo {author} {\bibfnamefont {P.}~\bibnamefont
  {Cadden-Zimansky}}, \bibinfo {author} {\bibfnamefont {Z.}~\bibnamefont
  {Jiang}}, \ and\ \bibinfo {author} {\bibfnamefont {P.}~\bibnamefont {Kim}},\
  }\href@noop {} {\bibfield  {journal} {\bibinfo  {journal} {Physical Review
  Letters},\ }\textbf {\bibinfo {volume} {104}},\ \bibinfo {pages} {066801}
  (\bibinfo {year} {2010})}\BibitemShut {NoStop}%
\bibitem [{\citenamefont {Nandkishore}\ and\ \citenamefont
  {Levitov}(2010){\natexlab{a}}}]{Nandkishore1}%
  \BibitemOpen
  \bibfield  {author} {\bibinfo {author} {\bibfnamefont {R.}~\bibnamefont
  {Nandkishore}}\ and\ \bibinfo {author} {\bibfnamefont {L.}~\bibnamefont
  {Levitov}},\ }\href@noop {} {\bibfield  {journal} {\bibinfo  {journal}
  {Physical Review Letters},\ }\textbf {\bibinfo {volume} {104}},\ \bibinfo
  {pages} {156803} (\bibinfo {year} {2010}{\natexlab{a}})}\BibitemShut
  {NoStop}%
\bibitem [{\citenamefont {Nandkishore}\ and\ \citenamefont
  {Levitov}(2010){\natexlab{b}}}]{Nandkishore2}%
  \BibitemOpen
  \bibfield  {author} {\bibinfo {author} {\bibfnamefont {R.}~\bibnamefont
  {Nandkishore}}\ and\ \bibinfo {author} {\bibfnamefont {L.}~\bibnamefont
  {Levitov}},\ }\href@noop {} {\bibfield  {journal} {\bibinfo  {journal}
  {Physical Review B},\ }\textbf {\bibinfo {volume} {82}},\ \bibinfo {pages}
  {115431} (\bibinfo {year} {2010}{\natexlab{b}})}\BibitemShut {NoStop}%
\bibitem [{\citenamefont {Zhang}\ \emph {et~al.}(2011)\citenamefont {Zhang},
  \citenamefont {Jung}, \citenamefont {Fiete}, \citenamefont {Niu},\ and\
  \citenamefont {MacDonald}}]{Zhang2}%
  \BibitemOpen
  \bibfield  {author} {\bibinfo {author} {\bibfnamefont {F.}~\bibnamefont
  {Zhang}}, \bibinfo {author} {\bibfnamefont {J.}~\bibnamefont {Jung}},
  \bibinfo {author} {\bibfnamefont {G.}~\bibnamefont {Fiete}}, \bibinfo
  {author} {\bibfnamefont {Q.}~\bibnamefont {Niu}}, \ and\ \bibinfo {author}
  {\bibfnamefont {A.~H.}\ \bibnamefont {MacDonald}},\ }\href@noop {} {\bibfield
   {journal} {\bibinfo  {journal} {arXiv:1010.4003v1}} (\bibinfo {year}
  {2011})}\BibitemShut {NoStop}%
\bibitem [{\citenamefont {Jung}\ \emph {et~al.}()\citenamefont {Jung},
  \citenamefont {Zhang},\ and\ \citenamefont {MacDonald}}]{Jung1}%
  \BibitemOpen
  \bibfield  {author} {\bibinfo {author} {\bibfnamefont {J.}~\bibnamefont
  {Jung}}, \bibinfo {author} {\bibfnamefont {F.}~\bibnamefont {Zhang}}, \ and\
  \bibinfo {author} {\bibfnamefont {A.~H.}\ \bibnamefont {MacDonald}},\
  }\href@noop {} {\bibfield  {journal} {\bibinfo  {journal} {Physical Review
  B},\ }\textbf {\bibinfo {volume} {83}},\ \bibinfo {pages}
  {115408}}\BibitemShut {NoStop}%
\bibitem [{\citenamefont {Martin}\ \emph {et~al.}()\citenamefont {Martin},
  \citenamefont {Feldman}, \citenamefont {Weitz}, \citenamefont {Allen},\ and\
  \citenamefont {Yacoby}}]{Martin1}%
  \BibitemOpen
  \bibfield  {author} {\bibinfo {author} {\bibfnamefont {J.}~\bibnamefont
  {Martin}}, \bibinfo {author} {\bibfnamefont {B.~E.}\ \bibnamefont {Feldman}},
  \bibinfo {author} {\bibfnamefont {R.~T.}\ \bibnamefont {Weitz}}, \bibinfo
  {author} {\bibfnamefont {M.~T.}\ \bibnamefont {Allen}}, \ and\ \bibinfo
  {author} {\bibfnamefont {A.}~\bibnamefont {Yacoby}},\ }\href@noop {}
  {\bibfield  {journal} {\bibinfo  {journal} {Physical Review Letters},\
  }\textbf {\bibinfo {volume} {105}},\ \bibinfo {pages} {256806}}\BibitemShut
  {NoStop}%
\bibitem [{\citenamefont {Barlas}\ \emph {et~al.}(2008)\citenamefont {Barlas},
  \citenamefont {Cote}, \citenamefont {Nomura},\ and\ \citenamefont
  {MacDonald}}]{Barlas1}%
  \BibitemOpen
  \bibfield  {author} {\bibinfo {author} {\bibfnamefont {Y.}~\bibnamefont
  {Barlas}}, \bibinfo {author} {\bibfnamefont {R.}~\bibnamefont {Cote}},
  \bibinfo {author} {\bibfnamefont {K.}~\bibnamefont {Nomura}}, \ and\ \bibinfo
  {author} {\bibfnamefont {A.~H.}\ \bibnamefont {MacDonald}},\ }\href@noop {}
  {\bibfield  {journal} {\bibinfo  {journal} {Physical Review Letters},\
  }\textbf {\bibinfo {volume} {101}},\ \bibinfo {pages} {097601} (\bibinfo
  {year} {2008})}\BibitemShut {NoStop}%
\bibitem [{\citenamefont {Tombros}\ \emph
  {et~al.}(2011){\natexlab{b}}\citenamefont {Tombros}, \citenamefont
  {Veligura}, \citenamefont {Junesch}, \citenamefont {van~den Berg},
  \citenamefont {Zomer}, \citenamefont {Wojtaszek}, \citenamefont {Marun},
  \citenamefont {Jonkman},\ and\ \citenamefont {van Wees}}]{Tombros1}%
  \BibitemOpen
  \bibfield  {author} {\bibinfo {author} {\bibfnamefont {N.}~\bibnamefont
  {Tombros}}, \bibinfo {author} {\bibfnamefont {A.}~\bibnamefont {Veligura}},
  \bibinfo {author} {\bibfnamefont {J.}~\bibnamefont {Junesch}}, \bibinfo
  {author} {\bibfnamefont {J.~J.}\ \bibnamefont {van~den Berg}}, \bibinfo
  {author} {\bibfnamefont {P.~J.}\ \bibnamefont {Zomer}}, \bibinfo {author}
  {\bibfnamefont {M.}~\bibnamefont {Wojtaszek}}, \bibinfo {author}
  {\bibfnamefont {I.~J.~V.}\ \bibnamefont {Marun}}, \bibinfo {author}
  {\bibfnamefont {H.~T.}\ \bibnamefont {Jonkman}}, \ and\ \bibinfo {author}
  {\bibfnamefont {B.~J.}\ \bibnamefont {van Wees}},\ }\href@noop {} {\bibfield
  {journal} {\bibinfo  {journal} {Journal of Applied Physics},\ }\textbf
  {\bibinfo {volume} {109}},\ \bibinfo {pages} {093702} (\bibinfo {year}
  {2011}{\natexlab{b}})}\BibitemShut {NoStop}%
\bibitem [{\citenamefont {Novoselov}\ \emph {et~al.}(2004)\citenamefont
  {Novoselov}, \citenamefont {Geim}, \citenamefont {Morozov}, \citenamefont
  {Jiang}, \citenamefont {Zhang}, \citenamefont {Dubonos}, \citenamefont
  {Grigorieva},\ and\ \citenamefont {Firsov}}]{Novoselov_firstgraphene}%
  \BibitemOpen
  \bibfield  {author} {\bibinfo {author} {\bibfnamefont {K.~S.}\ \bibnamefont
  {Novoselov}}, \bibinfo {author} {\bibfnamefont {A.~K.}\ \bibnamefont {Geim}},
  \bibinfo {author} {\bibfnamefont {S.~V.}\ \bibnamefont {Morozov}}, \bibinfo
  {author} {\bibfnamefont {D.}~\bibnamefont {Jiang}}, \bibinfo {author}
  {\bibfnamefont {Y.}~\bibnamefont {Zhang}}, \bibinfo {author} {\bibfnamefont
  {S.~V.}\ \bibnamefont {Dubonos}}, \bibinfo {author} {\bibfnamefont {I.~V.}\
  \bibnamefont {Grigorieva}}, \ and\ \bibinfo {author} {\bibfnamefont {A.~A.}\
  \bibnamefont {Firsov}},\ }\href@noop {} {\bibfield  {journal} {\bibinfo
  {journal} {Science},\ }\textbf {\bibinfo {volume} {306}},\ \bibinfo {pages}
  {666} (\bibinfo {year} {2004})}\BibitemShut {NoStop}%
\bibitem [{\citenamefont {Blake}\ \emph {et~al.}(2007)\citenamefont {Blake},
  \citenamefont {Hill}, \citenamefont {Neto}, \citenamefont {Novoselov},
  \citenamefont {Jiang}, \citenamefont {Yang}, \citenamefont {Booth},\ and\
  \citenamefont {Geim}}]{Blake1}%
  \BibitemOpen
  \bibfield  {author} {\bibinfo {author} {\bibfnamefont {P.}~\bibnamefont
  {Blake}}, \bibinfo {author} {\bibfnamefont {E.~W.}\ \bibnamefont {Hill}},
  \bibinfo {author} {\bibfnamefont {A.~H.~C.}\ \bibnamefont {Neto}}, \bibinfo
  {author} {\bibfnamefont {K.~S.}\ \bibnamefont {Novoselov}}, \bibinfo {author}
  {\bibfnamefont {D.}~\bibnamefont {Jiang}}, \bibinfo {author} {\bibfnamefont
  {R.}~\bibnamefont {Yang}}, \bibinfo {author} {\bibfnamefont {T.~J.}\
  \bibnamefont {Booth}}, \ and\ \bibinfo {author} {\bibfnamefont {A.~K.}\
  \bibnamefont {Geim}},\ }\href@noop {} {\bibfield  {journal} {\bibinfo
  {journal} {Applied Physics Letters},\ }\textbf {\bibinfo {volume} {91}},\
  \bibinfo {pages} {063124} (\bibinfo {year} {2007})}\BibitemShut {NoStop}%
\bibitem [{\citenamefont {Vera-Marun}\ \emph {et~al.}(2011)\citenamefont
  {Vera-Marun}, \citenamefont {Zomer}, \citenamefont {Veligura}, \citenamefont
  {Guimaraes}, \citenamefont {Visser}, \citenamefont {Tombros}, \citenamefont
  {van Elferen}, \citenamefont {Zeitler},\ and\ \citenamefont {van
  Wees}}]{Marun}%
  \BibitemOpen
  \bibfield  {author} {\bibinfo {author} {\bibfnamefont {I.~J.}\ \bibnamefont
  {Vera-Marun}}, \bibinfo {author} {\bibfnamefont {P.~J.}\ \bibnamefont
  {Zomer}}, \bibinfo {author} {\bibfnamefont {A.}~\bibnamefont {Veligura}},
  \bibinfo {author} {\bibfnamefont {M.~H.~D.}\ \bibnamefont {Guimaraes}},
  \bibinfo {author} {\bibfnamefont {L.}~\bibnamefont {Visser}}, \bibinfo
  {author} {\bibfnamefont {N.}~\bibnamefont {Tombros}}, \bibinfo {author}
  {\bibfnamefont {H.~J.}\ \bibnamefont {van Elferen}}, \bibinfo {author}
  {\bibfnamefont {U.}~\bibnamefont {Zeitler}}, \ and\ \bibinfo {author}
  {\bibfnamefont {B.~J.}\ \bibnamefont {van Wees}},\ }\href@noop {} {\bibfield
  {journal} {\bibinfo  {journal} {(submitted to arxive)}} (\bibinfo {year}
  {2011})}\BibitemShut {NoStop}%
\bibitem [{\citenamefont {Moser}\ \emph {et~al.}(2007)\citenamefont {Moser},
  \citenamefont {Barreiro},\ and\ \citenamefont {Bachtold}}]{Moser1}%
  \BibitemOpen
  \bibfield  {author} {\bibinfo {author} {\bibfnamefont {J.}~\bibnamefont
  {Moser}}, \bibinfo {author} {\bibfnamefont {A.}~\bibnamefont {Barreiro}}, \
  and\ \bibinfo {author} {\bibfnamefont {A.}~\bibnamefont {Bachtold}},\
  }\href@noop {} {\bibfield  {journal} {\bibinfo  {journal} {Applied Physics
  Letters},\ }\textbf {\bibinfo {volume} {91}},\ \bibinfo {pages} {163513}
  (\bibinfo {year} {2007})}\BibitemShut {NoStop}%
\bibitem [{\citenamefont {Abanin}\ and\ \citenamefont
  {Levitov}(2008)}]{Abanin1}%
  \BibitemOpen
  \bibfield  {author} {\bibinfo {author} {\bibfnamefont {D.~A.}\ \bibnamefont
  {Abanin}}\ and\ \bibinfo {author} {\bibfnamefont {L.~S.}\ \bibnamefont
  {Levitov}},\ }\href@noop {} {\bibfield  {journal} {\bibinfo  {journal}
  {Physical Review B},\ }\textbf {\bibinfo {volume} {78}},\ \bibinfo {pages}
  {035416} (\bibinfo {year} {2008})}\BibitemShut {NoStop}%
\bibitem [{\citenamefont {Huard}\ \emph {et~al.}(2008)\citenamefont {Huard},
  \citenamefont {Stander}, \citenamefont {Sulpizio},\ and\ \citenamefont
  {Goldhaber-Gordon}}]{Huard1}%
  \BibitemOpen
  \bibfield  {author} {\bibinfo {author} {\bibfnamefont {B.}~\bibnamefont
  {Huard}}, \bibinfo {author} {\bibfnamefont {N.}~\bibnamefont {Stander}},
  \bibinfo {author} {\bibfnamefont {J.~A.}\ \bibnamefont {Sulpizio}}, \ and\
  \bibinfo {author} {\bibfnamefont {D.}~\bibnamefont {Goldhaber-Gordon}},\
  }\href@noop {} {\bibfield  {journal} {\bibinfo  {journal} {Physical Review
  B},\ }\textbf {\bibinfo {volume} {78}},\ \bibinfo {pages} {121402} (\bibinfo
  {year} {2008})}\BibitemShut {NoStop}%
\bibitem [{\citenamefont {Blake}\ \emph {et~al.}(2009)\citenamefont {Blake},
  \citenamefont {Yang}, \citenamefont {Morozov}, \citenamefont {Schedin},
  \citenamefont {Ponomarenko}, \citenamefont {Zhukov}, \citenamefont {Nair},
  \citenamefont {Grigorieva}, \citenamefont {Novoselov},\ and\ \citenamefont
  {Geim}}]{Blake2}%
  \BibitemOpen
  \bibfield  {author} {\bibinfo {author} {\bibfnamefont {P.}~\bibnamefont
  {Blake}}, \bibinfo {author} {\bibfnamefont {R.}~\bibnamefont {Yang}},
  \bibinfo {author} {\bibfnamefont {S.~V.}\ \bibnamefont {Morozov}}, \bibinfo
  {author} {\bibfnamefont {F.}~\bibnamefont {Schedin}}, \bibinfo {author}
  {\bibfnamefont {L.~A.}\ \bibnamefont {Ponomarenko}}, \bibinfo {author}
  {\bibfnamefont {A.~A.}\ \bibnamefont {Zhukov}}, \bibinfo {author}
  {\bibfnamefont {R.~R.}\ \bibnamefont {Nair}}, \bibinfo {author}
  {\bibfnamefont {I.~V.}\ \bibnamefont {Grigorieva}}, \bibinfo {author}
  {\bibfnamefont {K.~S.}\ \bibnamefont {Novoselov}}, \ and\ \bibinfo {author}
  {\bibfnamefont {A.~K.}\ \bibnamefont {Geim}},\ }\href@noop {} {\bibfield
  {journal} {\bibinfo  {journal} {Solid State Communications},\ }\textbf
  {\bibinfo {volume} {149}},\ \bibinfo {pages} {1068} (\bibinfo {year}
  {2009})}\BibitemShut {NoStop}%
\bibitem [{\citenamefont {McCann}(2006)}]{Mccannbilayers}%
  \BibitemOpen
  \bibfield  {author} {\bibinfo {author} {\bibfnamefont {E.}~\bibnamefont
  {McCann}},\ }\href@noop {} {\bibfield  {journal} {\bibinfo  {journal}
  {Physical Review B},\ }\textbf {\bibinfo {volume} {74}},\ \bibinfo {pages}
  {161403} (\bibinfo {year} {2006})}\BibitemShut {NoStop}%
\bibitem [{\citenamefont {Kim}\ \emph {et~al.}(2011)\citenamefont {Kim},
  \citenamefont {Lee},\ and\ \citenamefont {Tutuc}}]{SeyoungKim1}%
  \BibitemOpen
  \bibfield  {author} {\bibinfo {author} {\bibfnamefont {S.}~\bibnamefont
  {Kim}}, \bibinfo {author} {\bibfnamefont {K.}~\bibnamefont {Lee}}, \ and\
  \bibinfo {author} {\bibfnamefont {E.}~\bibnamefont {Tutuc}},\ }\href@noop {}
  {\bibfield  {journal} {\bibinfo  {journal} {Physical Review Letters},\
  }\textbf {\bibinfo {volume} {107}},\ \bibinfo {pages} {016803} (\bibinfo
  {year} {2011})}\BibitemShut {NoStop}%
\bibitem [{\citenamefont {Veligura}\ \emph {et~al.}(2012)\citenamefont
  {Veligura}, \citenamefont {van Elferen}, \citenamefont {Tombros},
  \citenamefont {Maan}, \citenamefont {Zeitler},\ and\ \citenamefont {van
  Wees}}]{Alina1}%
  \BibitemOpen
  \bibfield  {author} {\bibinfo {author} {\bibfnamefont {A.}~\bibnamefont
  {Veligura}}, \bibinfo {author} {\bibfnamefont {H.J.}~\bibnamefont {van
  Elferen}}, \bibinfo {author} {\bibfnamefont {N.}~\bibnamefont {Tombros}},
  \bibinfo {author} {\bibfnamefont {J.~C.}~\bibnamefont {Maan}}, \bibinfo {author}
  {\bibfnamefont {U.}~\bibnamefont {Zeitler}}, \ and\ \bibinfo {author}
  {\bibfnamefont {B.~J.}~\bibnamefont {van Wees}},\ }\href@noop {} {\bibfield
  {journal} {\bibinfo  {journal} {arXiv:1202.1753}} (\bibinfo {year}
  {2012})}\BibitemShut {NoStop}%
\bibitem [{\citenamefont {Kurganova}\ \emph {et~al.}(2011)\citenamefont
  {Kurganova}, \citenamefont {van Elferen}, \citenamefont {McCollam},
  \citenamefont {Ponomarenko}, \citenamefont {Novoselov}, \citenamefont
  {Veligura}, \citenamefont {van Wees}, \citenamefont {Maan},\ and\
  \citenamefont {Zeitler}}]{Kurganova1}%
  \BibitemOpen
  \bibfield  {author} {\bibinfo {author} {\bibfnamefont {E.~V.}\ \bibnamefont
  {Kurganova}}, \bibinfo {author} {\bibfnamefont {H.~J.}\ \bibnamefont {van
  Elferen}}, \bibinfo {author} {\bibfnamefont {A.}~\bibnamefont {McCollam}},
  \bibinfo {author} {\bibfnamefont {L.~A.}\ \bibnamefont {Ponomarenko}},
  \bibinfo {author} {\bibfnamefont {K.~S.}\ \bibnamefont {Novoselov}}, \bibinfo
  {author} {\bibfnamefont {A.}~\bibnamefont {Veligura}}, \bibinfo {author}
  {\bibfnamefont {B.~J.}\ \bibnamefont {van Wees}}, \bibinfo {author}
  {\bibfnamefont {J.~C.}\ \bibnamefont {Maan}}, \ and\ \bibinfo {author}
  {\bibfnamefont {U.}~\bibnamefont {Zeitler}},\ }\href@noop {} {\bibfield
  {journal} {\bibinfo  {journal} {Physical Review B},\ }\textbf {\bibinfo
  {volume} {84}},\ \bibinfo {pages} {121407} (\bibinfo {year}
  {2011})}\BibitemShut {NoStop}%
\bibitem [{\citenamefont {Wang}\ \emph {et~al.}(2010)\citenamefont {Wang},
  \citenamefont {Philippe},\ and\ \citenamefont {Elias}}]{Wang1}%
  \BibitemOpen
  \bibfield  {author} {\bibinfo {author} {\bibfnamefont {Z.}~\bibnamefont
  {Wang}}, \bibinfo {author} {\bibfnamefont {L.}~\bibnamefont {Philippe}}, \
  and\ \bibinfo {author} {\bibfnamefont {J.}~\bibnamefont {Elias}},\
  }\href@noop {} {\bibfield  {journal} {\bibinfo  {journal} {Physical Review
  B},\ }\textbf {\bibinfo {volume} {81}},\ \bibinfo {pages} {155405} (\bibinfo
  {year} {2010})}\BibitemShut {NoStop}%
\bibitem [{\citenamefont {Stormer}\ \emph {et~al.}(1992)\citenamefont
  {Stormer}, \citenamefont {Baldwin}, \citenamefont {Pfeiffer},\ and\
  \citenamefont {West}}]{Stormer1}%
  \BibitemOpen
  \bibfield  {author} {\bibinfo {author} {\bibfnamefont {H.~L.}\ \bibnamefont
  {Stormer}}, \bibinfo {author} {\bibfnamefont {K.~W.}\ \bibnamefont
  {Baldwin}}, \bibinfo {author} {\bibfnamefont {L.~N.}\ \bibnamefont
  {Pfeiffer}}, \ and\ \bibinfo {author} {\bibfnamefont {K.~W.}\ \bibnamefont
  {West}},\ }\href@noop {} {\bibfield  {journal} {\bibinfo  {journal} {Solid
  State Communications},\ }\textbf {\bibinfo {volume} {84}},\ \bibinfo {pages}
  {95} (\bibinfo {year} {1992})}\BibitemShut {NoStop}%
\bibitem [{\citenamefont {Lifshitz}\ and\ \citenamefont
  {Kosevich}(1956)}]{Kosevic}%
  \BibitemOpen
  \bibfield  {author} {\bibinfo {author} {\bibfnamefont {I.~M.}\ \bibnamefont
  {Lifshitz}}\ and\ \bibinfo {author} {\bibfnamefont {A.~M.}\ \bibnamefont
  {Kosevich}},\ }\href@noop {} {\bibfield  {journal} {\bibinfo  {journal}
  {Soviet Physics Jetp-Ussr},\ }\textbf {\bibinfo {volume} {2}},\ \bibinfo
  {pages} {636} (\bibinfo {year} {1956})}\BibitemShut {NoStop}%
\bibitem [{\citenamefont {Kurganova}\ \emph {et~al.}(2010)\citenamefont
  {Kurganova}, \citenamefont {Giesbers}, \citenamefont {Gorbachev},
  \citenamefont {Geim}, \citenamefont {Novoselov}, \citenamefont {Maan},\ and\
  \citenamefont {Zeitler}}]{Kurganova2}%
  \BibitemOpen
  \bibfield  {author} {\bibinfo {author} {\bibfnamefont {E.~V.}\ \bibnamefont
  {Kurganova}}, \bibinfo {author} {\bibfnamefont {A.~J.~M.}\ \bibnamefont
  {Giesbers}}, \bibinfo {author} {\bibfnamefont {R.~V.}\ \bibnamefont
  {Gorbachev}}, \bibinfo {author} {\bibfnamefont {A.~K.}\ \bibnamefont {Geim}},
  \bibinfo {author} {\bibfnamefont {K.~S.}\ \bibnamefont {Novoselov}}, \bibinfo
  {author} {\bibfnamefont {J.~C.}\ \bibnamefont {Maan}}, \ and\ \bibinfo
  {author} {\bibfnamefont {U.}~\bibnamefont {Zeitler}},\ }\href@noop {}
  {\bibfield  {journal} {\bibinfo  {journal} {Solid State Communications},\
  }\textbf {\bibinfo {volume} {150}},\ \bibinfo {pages} {2209} (\bibinfo {year}
  {2010})}\BibitemShut {NoStop}%
\bibitem [{\citenamefont {Freitag}\ \emph {et~al.}(2011)\citenamefont
  {Freitag}, \citenamefont {Trbovic}, \citenamefont {Weiss},\ and\
  \citenamefont {Schönenberger}}]{Freitag}%
  \BibitemOpen
  \bibfield  {author} {\bibinfo {author} {\bibfnamefont {F.}~\bibnamefont
  {Freitag}}, \bibinfo {author} {\bibfnamefont {J.}~\bibnamefont {Trbovic}},
  \bibinfo {author} {\bibfnamefont {M.}~\bibnamefont {Weiss}}, \ and\ \bibinfo
  {author} {\bibfnamefont {C.}~\bibnamefont {Schönenberger}},\ }\href@noop {}
  {\bibfield  {journal} {\bibinfo  {journal} {arXiv:1104.3816v2}} (\bibinfo
  {year} {2011})}\BibitemShut {NoStop}%
\end{thebibliography}

%

\end{document}